\documentclass[letterpaper]{article}
\pdfoutput=1
\usepackage{dcolumn}
\usepackage{bm}

\usepackage{graphicx}
\usepackage{amssymb,amsmath}
\usepackage{multirow}
\usepackage{multicol}
\usepackage{nonfloat}
\usepackage{cite,color,url}
\usepackage[colorlinks=true
,urlcolor=blue
,anchorcolor=blue
,citecolor=blue
,filecolor=blue
,linkcolor=blue
,menucolor=blue
,linktocpage=true
,pdfproducer=medialab
,pdfa=true
]{hyperref}

\usepackage{epsfig,psfrag,rotating,soul}
\usepackage{rotfloat}

\evensidemargin \oddsidemargin
\allowdisplaybreaks

\psfrag{lk}{$l_k$}
\psfrag{lm}{$l_m$}
\psfrag{blk}{$\bar {l}_k$}
\psfrag{blm}{$\bar {l}_m$}

\def\thefootnote{\fnsymbol{footnote}}

\let\OLDthebibliography\thebibliography
\renewcommand\thebibliography[1]{
  \OLDthebibliography{#1}
  \setlength{\parskip}{0pt}
  \setlength{\itemsep}{0pt plus 0.3ex}
}

\newcommand\myfigure[1]{%
\medskip\noindent\begin{minipage}{\columnwidth}
\centering%
#1%
\end{minipage}\medskip}

\begin{document}

\begin{flushright}
IFT-UAM/CSIC-15-084\\
FTUAM-15-24\\
SU-HET-08-2015\\
\end{flushright}

\vspace{0.5cm}

\begin{center}

\begin{Large}
\textbf{\textsc{Exotic $\mu\tau j j$ events  from heavy ISS neutrinos at the LHC}}
\end{Large}

\vspace{1cm}

{\sc
E. Arganda$^{1}$%
\footnote{\tt \href{mailto:ernesto.arganda@unizar.es}{ernesto.arganda@unizar.es}}%
, M.J. Herrero$^{2}$%
\footnote{\tt \href{mailto:maria.herrero@uam.es}{maria.herrero@uam.es}}%
, X. Marcano$^{2}$%
\footnote{\tt \href{mailto:xabier.marcano@uam.es}{xabier.marcano@uam.es}}%
, C. Weiland$^{2,3}$%
\footnote{\tt \href{mailto:cedric.weiland@uam.es}{cedric.weiland@uam.es}}%
}

\vspace*{.7cm}

{\sl
$^1$Departamento de F\'{\i}sica Te\'orica, Facultad de Ciencias,\\
Universidad de Zaragoza, E-50009 Zaragoza, Spain

\vspace*{0.1cm}

$^2$Departamento de F\'{\i}sica Te\'orica and Instituto de F\'{\i}sica Te\'orica, IFT-UAM/CSIC,\\
Universidad Aut\'onoma de Madrid, Cantoblanco, 28049 Madrid, Spain

\vspace*{0.1cm}

$^3$Graduate School of Science and Engineering,\\
Shimane University, Matsue, 690-8504 Japan 
}

\end{center}

\vspace*{0.1cm}

\begin{abstract}
\noindent
In this letter we study new relevant phenomenological consequences of the right-handed heavy neutrinos with masses at the ${\cal O}(1)$ TeV energy scale, working  within the context of the Inverse Seesaw Model that includes three pairs of quasi-degenerate pseudo-Dirac heavy neutrinos.  We propose a new exotic signal of these heavy neutrinos at the CERN Large Hadron Collider containing a muon, a tau lepton, and two jets in the final state,  which is based on the interesting fact that this model can incorporate large Lepton Flavor Violation for specific choices of the relevant parameters, particularly, the neutrino Yukawa couplings. We will show here that an observable number of $\mu\tau jj$ exotic events, without missing energy, can be produced at this ongoing run of the LHC.
\end{abstract}

\vspace*{0.5cm}

\def\thefootnote{\arabic{footnote}}
\setcounter{footnote}{0}

\begin{multicols}{2}

\section{Introduction}
\label{intro}
One of the most interesting properties of the Inverse Seesaw Model (ISS) is that it allows for the existence of heavy Majorana neutrinos with masses within the reach of the CERN Large Hadron Collider (LHC), and with couplings to the Standard Model (SM) particles that can be sizable, therefore leading to observable production and decay rates. 
The version of the ISS that incorporates three pairs of SM-singlet heavy neutrinos, one pair per generation, is especially appealing because, on the one hand,  it can explain the observed small neutrino masses and neutrino oscillation data and, on the other hand in connection with the LHC physics, it can lead to new interesting  phenomenology with experimental signatures distinct from those of the SM particles.  
The most frequently studied signatures of  heavy neutrinos are those related to their Majorana nature\cite{PilaftsisDilepton,delAguila:2007em,Atre:2009rg} and, in particular, the most characteristic signal is the same-sign dilepton plus two jets events which is being searched for at the LHC. 
The rates of this type of events within the ISS are directly related to the Majorana mass matrix $\mu_X$ that introduces the Lepton Number Violation in the model, which in turn is built in as to precisely fit the observed light neutrino masses via a low scale realization of the Seesaw Mechanism where $M_{\mathrm{light}} \simeq m_D M_R^{T-1} \mu_X M_R^{-1} m_D^T\,$ and $m_D$ and $M_R$ are the Dirac and  extended right handed (RH) neutrino sector mass matrices respectively. 
The usually assumed hierarchy among the three input scales,  $\mu_X \ll m_D \ll M_R$, then produces the interesting mentioned features of the ISS. Furthermore, due to the smallness of $\mu_X$, the heavy neutrinos combine in nearly degenerate pairs to form pseudo-Dirac fermions with masses close to $M_R$ 
 and with a splitting of order $\mathcal O(\mu_X)$. This pseudo-Dirac character of the heavy neutrinos has also been explored in the literature in connection with the appearance of other interesting signals at the LHC, like the trilepton final state \cite{OkadaTrilepton,BambhaniyaTrilepton}, and other multi-lepton signals \cite{delAguila:2008cj}.
 
 We propose here a new exotic signal of the ISS heavy neutrinos at the LHC that is based on another interesting feature of the ISS, the fact that it can incorporate large Lepton Flavor Violation (LFV) for specific choices of the model parameters, particularly the Yukawa couplings $Y_\nu$, and lead to a new rich phenomenology, including the appearance of rare processes like the radiative decays, $l_i \to l_j \gamma \; (i \neq j)$ and others. 
Moreover, it has been shown recently \cite{Arganda:2014dta} that sizable branching ratios for the LFV Higgs boson decays, like $H \to \mu \tau$, can also be produced within this ISS context,  that can be enhanced in its supersymmetric version~\cite{Arganda:2015naa} for specific choices of the model parameters, reaching values which are compatible with the present excess observed by CMS and ATLAS in $H \to \mu \tau$~\cite{Khachatryan:2015kon,Aad:2015gha}, while respecting the present upper bounds on the related radiative decay $\tau \to \mu \gamma$~\cite{Aubert:2009ag}. 
Our specific proposal here is to look at rare LHC events with one muon, one tau lepton and two jets in the final state that are produced in these ISS scenarios with large LFV, and that presumably will have a very small SM background. This letter then summarizes our computation of the rates for these exotic $\mu \tau j j$ events due to the production and decays of the heavy quasi-Dirac neutrinos at the LHC within the ISS.

\section{ISS scenarios with large LFV}
\label{model}
We consider the ISS model that extends the SM with two kinds of  fermionic gauge singlets: RH neutrinos, denoted here by $\nu_R$ and coupled to the $\nu_L$ via Yukawa interactions,  and $X$, with opposite lepton number. We assume one pair of them per generation. The relevant ISS Lagrangian for the mass terms is:
\begin{equation}
 \label{LagrangianISS}
 \mathcal{L}_\mathrm{ISS} = - Y^{ij}_\nu \overline{L_{i}} \widetilde{\Phi} \nu_{Rj} - M_R^{ij} \overline{\nu_{Ri}^C} X_j - \frac{1}{2} \mu_{X}^{ij} \overline{X_{i}^C} X_{j} + \text{h.c.}\,,
\end{equation}
where $L$ is the SM lepton doublet, $\Phi$ is the SM Higgs doublet  with $\langle \Phi\rangle=v = 174\,\mathrm{GeV}$, $\widetilde{\Phi}=\imath \sigma_2 \Phi^*$,   $Y_\nu$ is the $3\times3$ neutrino Yukawa coupling matrix, $M_R$ is a lepton number conserving complex $3\times3$ mass matrix, and $\mu_X$ is
a Majorana complex $3\times3$ symmetric mass matrix that violates lepton number conservation by two units. 
The diagonalization of the relevant $9\times 9$ neutrino mass matrix, containing the Dirac mass matrix, $m_D=Y_\nu \langle \Phi\rangle$, and $M_R$,
\begin{equation}
\label{ISSmatrix}
 M_{\mathrm{ISS}}=\left(\begin{array}{c c c} 0 & m_D & 0 \\ m_D^T & 0 & M_R \\ 0 & M_R^T & \mu_X \end{array}\right)\,,
\end{equation}
provides the 9 physical neutrino states $n_j (j=1,..9)$: three light neutrinos $\nu_{1,2,3}$ that are identified with the ones experimentally observed, and six heavy neutrinos $N_{1,..6}$ which are the object of our interest here. 
In order to match the predictions of this ISS model with the low energy neutrino data, one considers small $\mu_X$  given by the following parameterization in terms of the light neutrino mixing matrix $U_{\rm PMNS}$\cite{PMNS}:
 \begin{equation}
\mu_X=M_R^T ~m_D^{-1}~ U_{\rm PMNS}^* m_\nu U_{\rm PMNS}^\dagger~ {m_D^T}^{-1} M_R.
\end{equation}
 This parametrization, which works quite well in the $\mu_X \ll m_D \ll M_R$ limit, was introduced in ref.\cite{Arganda:2014dta} and allows us to use $Y_\nu$ and $M_R$ as input parameters of the model. For the rest of this letter, we focus on the simplest case with diagonal $M_R$ matrix and degenerate entries, $M_{R_{1,2,3}} \equiv M_R$.  Within this mentioned small $\mu_X$ limit, the heavy neutrinos can be grouped into pseudo-Dirac pairs with nearly degenerate masses. Both the light neutrino masses and the small differences among the quasi-degenerate heavy neutrino masses are governed by the small size of $\mu_X$, basically following the same pattern as in the one generation case where: 
\begin{align}
 m_\nu &= \frac{m_{D}^2}{m_{D}^2+M_{R}^2} \mu_X\,\label{mnu},\\
 m_{N_{a/b}} &= \pm \sqrt{M_{R}^2+m_{D}^2} + \frac{M_{R}^2 \mu_X}{2 (m_{D}^2+M_{R}^2)}~.\label{mN}
\end{align}
For the present work, the corresponding pairings of the six heavy neutrinos are given by: 
${N_{1/2}}$, ${N_{3/4}}$ and  ${N_{5/6}}$, with $m_{N_{1/2}} \leq m_{N_{3/4}} \leq m_{N_{5/6}}$. Their corresponding masses in the $M_R\gg m_D$ limit are close to $M_R$ which will be taken here in the range accessible to the LHC, namely $\leq \mathcal O(1 \,{\rm TeV})$.

Regarding the heavy neutrino interactions, we give here just the relevant ones, namely, those to $W$ gauge bosons, which enter both in the heavy neutrino production and in the decays of our interest, and those to $Z$ gauge and Higgs bosons, which enter in the computation of the total heavy neutrino width:  
\begin{align}
\mathcal{L}_{W} &= -\frac{g}{\sqrt{2}} W^{\mu -}\bar{l_i} B_{l_i n_j} \gamma_{\mu} P_L n_j + \text{h.c}\,,\\ 
\mathcal{L}_Z &= -\frac g{4c_W}~Z^\mu \overline n_i \gamma_\mu  \big[ C_{n_i n_j} P_L -
C_{n_i n_j}^* P_R\big] n_j, \\
\mathcal{L}_{H} &= -\frac{g}{2 m_W} H \bar n_i C_{n_i n_j} \left[ m_{n_i} P_L + m_{n_j} P_R \right] n_j\,, 
\end{align}
where, assuming the charged leptons to be diagonal, %
\begin{eqnarray}
B_{l_i n_j} = U_{ij}^{\nu *}\,,\\
C_{n_i n_j} = \sum_{k=1}^3 U_{k i}^{\nu} U_{kj}^{\nu *}\ ,
\end{eqnarray}
and $U^\nu$ is the $9 \times 9$ neutrino rotation matrix that diagonalizes the mass matrix (\ref{ISSmatrix}) according to $U^{\nu T} M_{\mathrm{ISS}} U^\nu = \text{diag}(m_{n_1},\dots,m_{n_9})$.

Since we are interested in the ISS with large LFV, and  more specifically producing large $\tau-\mu$ transitions, we will work with some benchmark scenarios that produce these wanted features. In particular we  will use the three benchmark scenarios proposed in \cite{Arganda:2014dta} that are defined by the following input Yukawa matrices, $Y_\nu$:

\begin{subequations}
\begin{equation}
Y_{\tau \mu}^{(1)}=f \left(\begin{array}{ccc}
0&1&-1\\0.9&1&1\\1&1&1
\end{array}\right)~,~
\label{Ytmmax1}
\end{equation}
\begin{equation}
Y_{\tau \mu}^{(2)}=f \left(\begin{array}{ccc}
0&1&1\\1&1&-1\\-1&1&-1
\end{array}\right)~,~
\label{Ytmmax2}
\end{equation}
\begin{equation}
Y_{\tau \mu}^{(3)}=f \left(\begin{array}{ccc}
0&-1&1\\-1&1&1\\0.8&0.5&0.5
\end{array}\right).
\label{Ytmmax3}
\end{equation}
\label{Ytmmax}
\end{subequations}

These scenarios have been explored fully in \cite{Arganda:2014dta} and display large LFV ratios  for processes involving $\tau-\mu$ transitions (this motivates the given name, $Y_\nu=Y_{\tau \mu}^{(1,2,3)}$), whereas they strongly suppress processes involving  $\tau-e$ and 
$\mu-e$ transitions,  which can be easily understood with the use of the approximate formulas presented also in  \cite{Arganda:2014dta}:  
\begin{equation}
{\rm BR}^{\rm approx}_{\tau \to \mu \gamma}=8 \times 10^{-17} \frac{m_{\tau}^5({\rm GeV}^5)}{\Gamma_{\tau}{\rm (GeV)}} 
\bigg|\frac{v^2}{2M_R^2}(Y_\nu Y_\nu^\dagger)_{23}\bigg|^2 \,,
\label{approxformula} 
\end{equation} 
\begin{equation}\label{FIThtaumu}
{\rm BR}^{\rm approx}_{H\to\mu\bar\tau}=10^{-7}\frac{v^4}{M_R^4}~\Big|(Y_\nu Y_\nu^\dagger)_{23}-5.7(Y_\nu Y_\nu^\dagger Y_\nu Y_\nu^\dagger)_{23}\Big|^2 \,.
\end{equation}
For $f$ values close to its perturbativity limit, $f~=~\sqrt{6\pi}$, and $M_R \sim \mathcal O(1\,\rm{TeV})$ these lead to rates of up to ${\rm BR} {(H\to\mu\bar\tau)}\sim 10^{-5}$  while respecting the present bounds on the radiative decays,  ${\rm BR}(\mu\to~e\gamma)\leq 5.7\times 10^{-13}$\cite{Adam:2013mnn}, ${\rm BR}(\tau\to~\mu\gamma)\leq~4.4\times10^{-8}$ and  ${\rm BR}(\tau\to e\gamma)\leq 3.3\times 10^{-8}$\cite{Aubert:2009ag}. Moreover, in the case of the SUSY version of this ISS model, the Higgs decay rates induced by the SUSY loop contributions have been found in \cite{Arganda:2015naa} to be as large as the present LHC sensitivity of about ${\rm BR} ({H\to\mu \tau}) \sim 10^{-2}$~\cite{Khachatryan:2015kon,Aad:2015gha}.

As we see from eq.~(\ref{approxformula}), the constrains from LFV radiative decays set upper bounds on the combination $|(Y_\nu^{} Y_\nu^\dagger)_{ij}|$, with $i\neq j$. In terms of the heavy neutrino flavor mixing matrix, $B_{lN}$, they constrain the combination {$|(B_{lN}^{} B_{lN}^\dag)_{ij}|$}, but not the {$B_{l_iN_j}^{}$} itself, which is the relevant element that controls the flavor pattern of each of the heavy neutrinos.

\myfigure{
\includegraphics[width=.6\columnwidth]{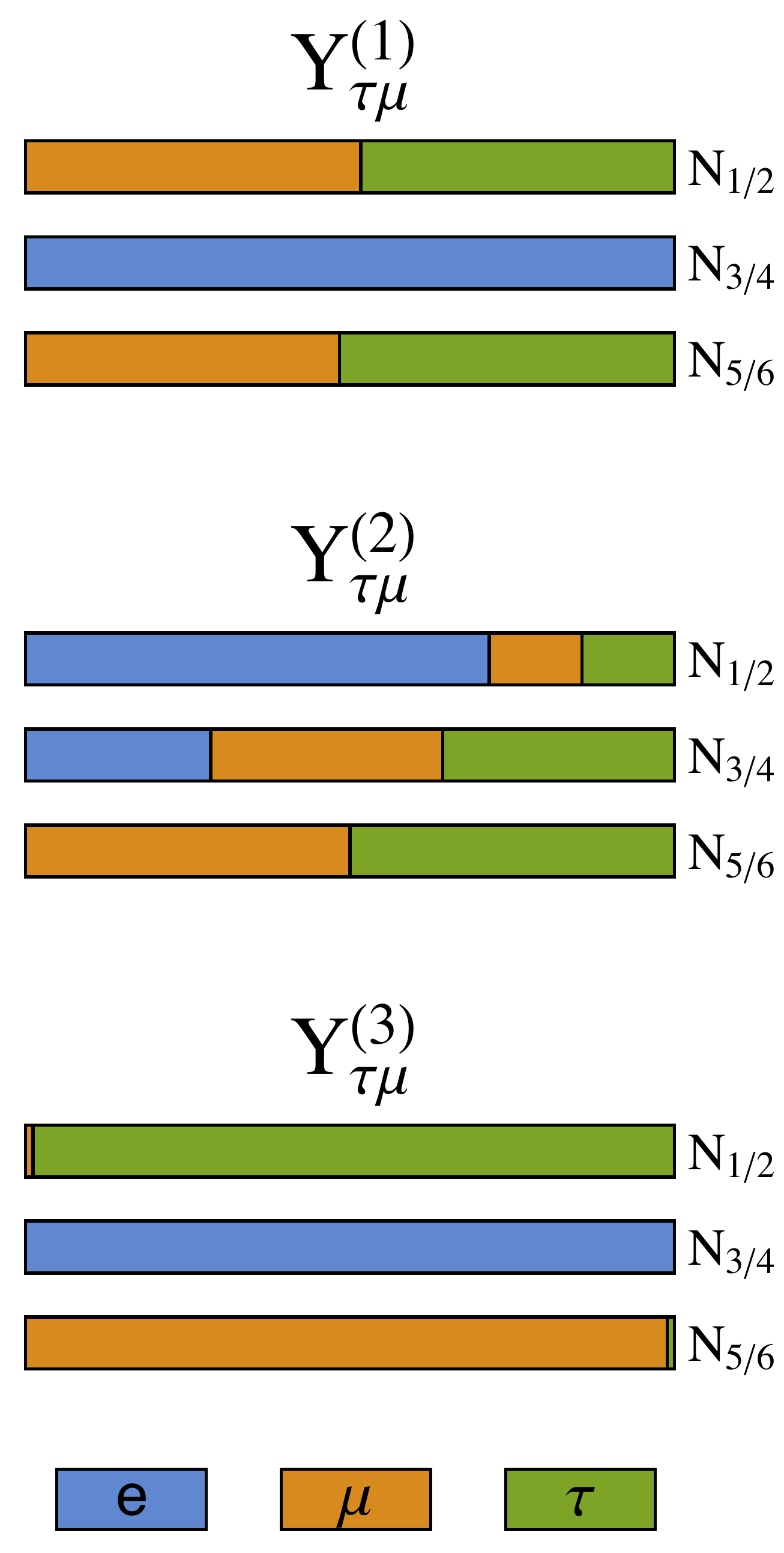}
\figcaption{Heavy neutrino flavor mixing, according to eq.(\ref{mixeq}), for the three benchmark scenarios in (\ref{Ytmmax}).}
\label{mixings}}

In figure \ref{mixings} we show the flavor content of each heavy neutrino for the three benchmark scenarios in (\ref{Ytmmax}), where the length of the colored bars is calculated as 
\begin{equation}
S_{lN_i}= \dfrac{\displaystyle{|B_{lN_i}|^2}}{\displaystyle{\sum_{l=e,\mu,\tau} |B_{lN_i}|^2}} \,,
\label{mixeq}
\end{equation}
and, therefore, represents the relative mixing of the heavy neutrino $N_i$ with a given flavor $l$.
We learn from fig.\ref{mixings} that, although the three scenarios share the property of suppressing the LFV $\mu$-$e$ and $\tau$-$e$ rates while maximizing the $\tau$-$\mu$ ones,  the heavy neutrino flavor mixing pattern is different in each scenario.
We also see that some heavy neutrinos carry both $\mu$ and $\tau$ flavors, specially in the first two scenarios, pointing towards signals with both $\mu$ and $\tau$ leptons. 
It should be further noticed, that the values of these ${B_{lN}}$ mixing parameters are determined within the ISS in terms of the input $m_D$ and $M_R$ mass matrices and, therefore, in the limit $m_D \ll M_R$ they are suppressed as ${B_{lN}} \sim \mathcal O (m_DM_R^{-1})$.  
In addition, they are constrained from ElectroWeak Precision Observables (EWPO) \cite{delAguila:2008pw} by: 
\begin{align}
|{B_{eN}}|^2&<3.0\times10^{-3}\,,\\
|{B_{\mu N}}|^2&<3.2\times10^{-3}\,,\\
|{B_{\tau N}}|^2&<6.2\times10^{-3}\,.
\end{align} 
\section{Numerical Results}
\label{results}

Heavy neutrinos with masses of the TeV order can be produced at present and future colliders, in particular in the second run of the LHC. 
The dominant production mechanism in this case is the Drell-Yan (DY)  process, fig.(\ref{FDs}) left, where the heavy neutrino is produced in association with a charged lepton.
The $\gamma W$ fusion, fig.(\ref{FDs}) right, also produces the same signal with two extra jets and, in fact, can  be relevant especially for large neutrino masses  in the $\mathcal O(1 ~\rm{TeV})$ energy range\cite{Alva:2014gxa,Dev:2013wba}.

In order to estimate the heavy neutrino production at the LHC, we have implemented the model in MadGraph 5\cite{Alwall:2014hca} using FeynRules\cite{Alloul:2013bka}. 
Following ref.\cite{Alva:2014gxa}, we have used a $K$-factor of $1.2$ for the DY-process and split the $\gamma W$ process into three regimes characterized by the virtuality of the photon\footnote{We warmly thank Richard Ruiz, Tao Han and Daniel Alva for their generous help  and clarifications in the implementation of the $\gamma W$ process.}: elastic, inelastic and deep inelastic scattering (DIS) regimes. In particular, we have set the boundaries between these three regimes to $\Lambda_\gamma^{\rm Elas}=1.22$~GeV and $\Lambda_\gamma^{\rm DIS}=15$~GeV.
In order to regularize possible collinear singularities, we have also imposed the following cuts to the transverse momentum and pseudorapidity of the outgoing leptons:
\begin{equation}
p_T^l > 10~ {\rm GeV}, \quad |\eta^l|<2.4~.
\end{equation} 

The results for the three benchmark scenarios in (\ref{Ytmmax}) are shown in fig.~\ref{production}, where the dominant DY production cross sections normalized by $f^2$ are plotted as functions of the mass parameter $M_R$.
We see that the production cross sections can be of the fb order, reachable at the LHC, for masses $M_R\lesssim600$ GeV.
Notice that the results are always equal for the pseudo-Dirac pairs, since their Majorana character plays a subleading role in their production.

\myfigure{
\includegraphics[width=.49\columnwidth]{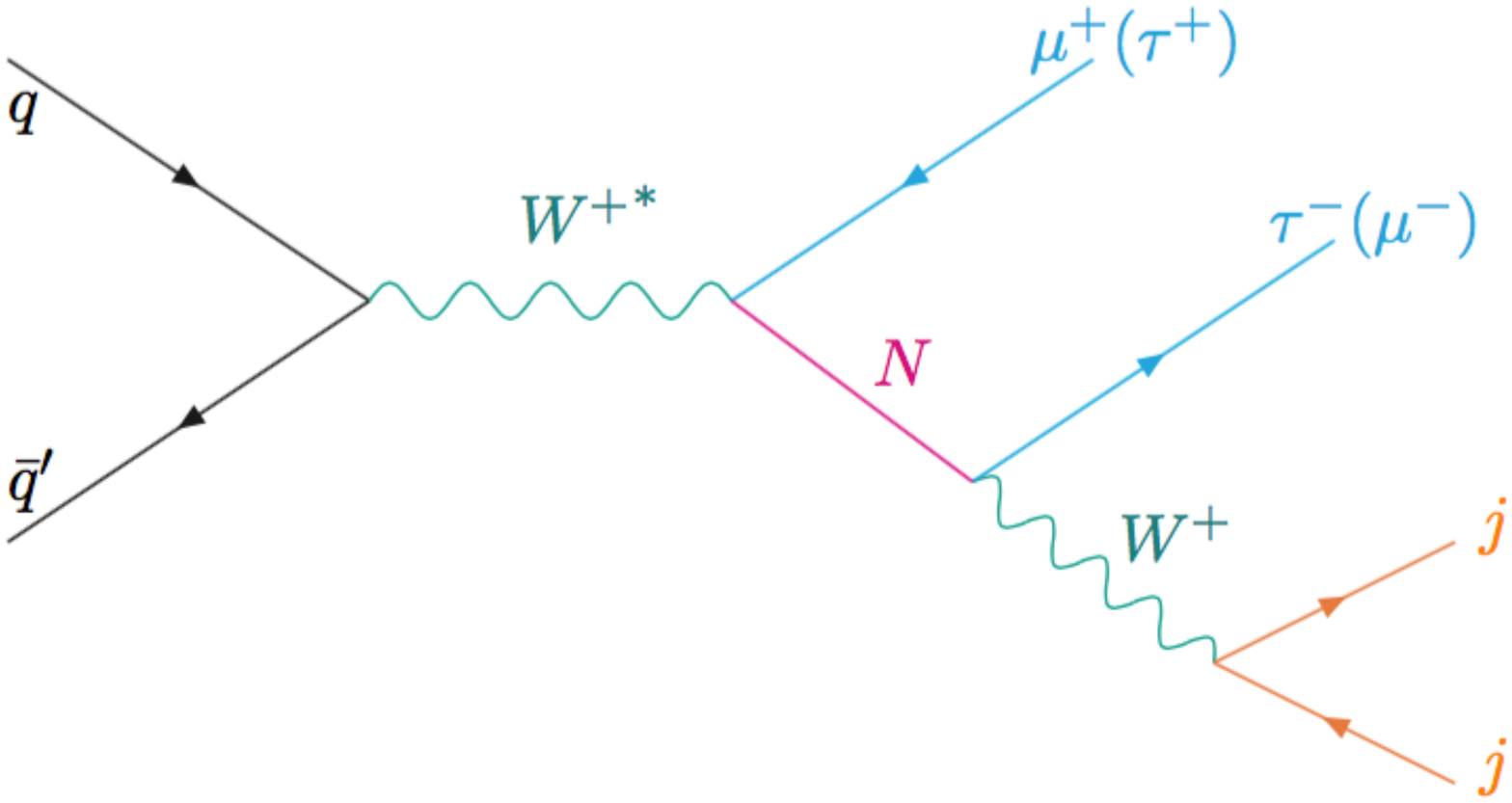}
\includegraphics[width=.49\columnwidth]{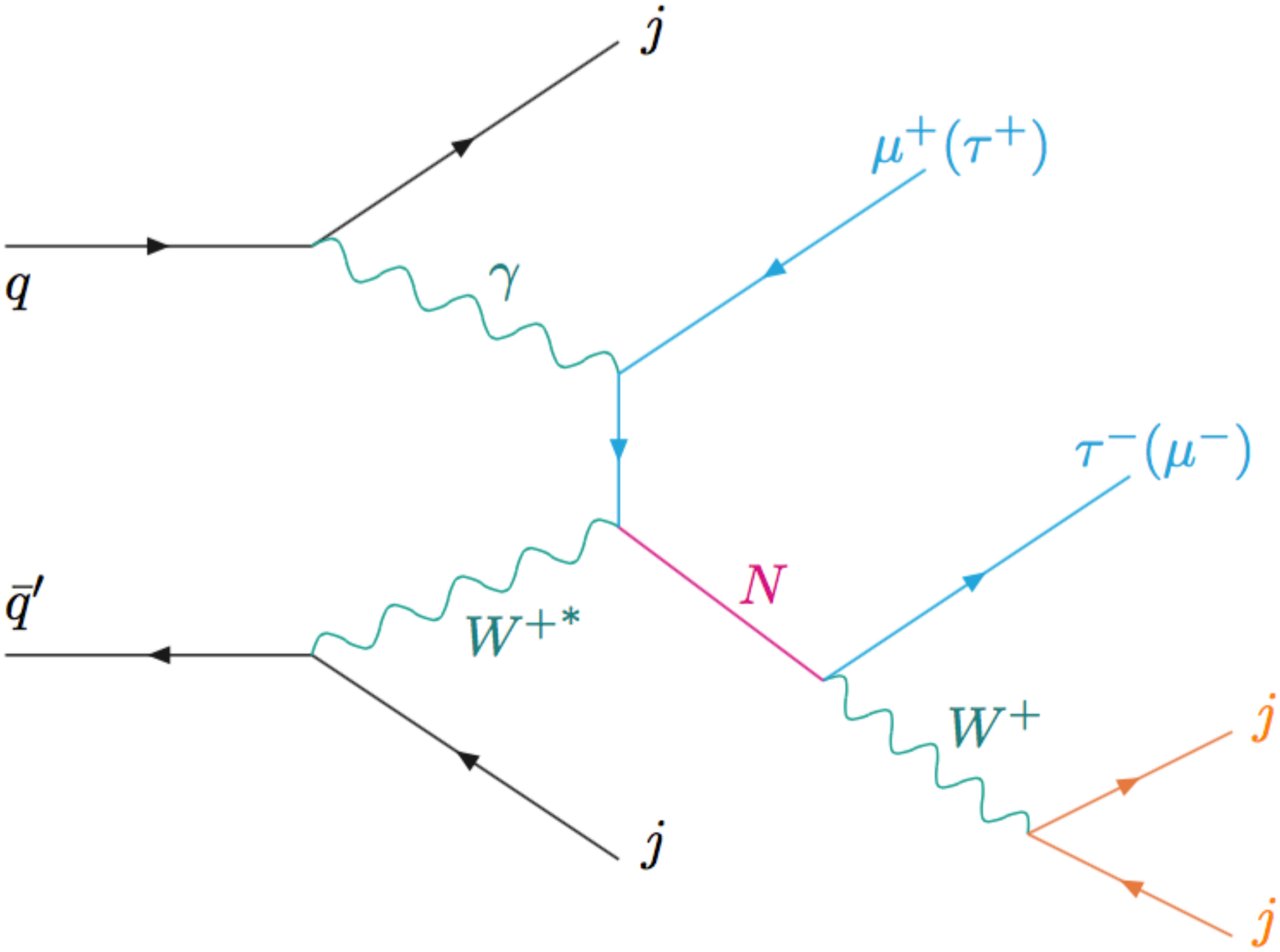}
\figcaption{The two main  processes, Drell-Yan and $\gamma W$ fusion, producing exotic $\tau \mu j j$ events via heavy neutrino production and decay at the LHC.}
\label{FDs}}
~

\myfigure{
\includegraphics[width=.95\columnwidth]{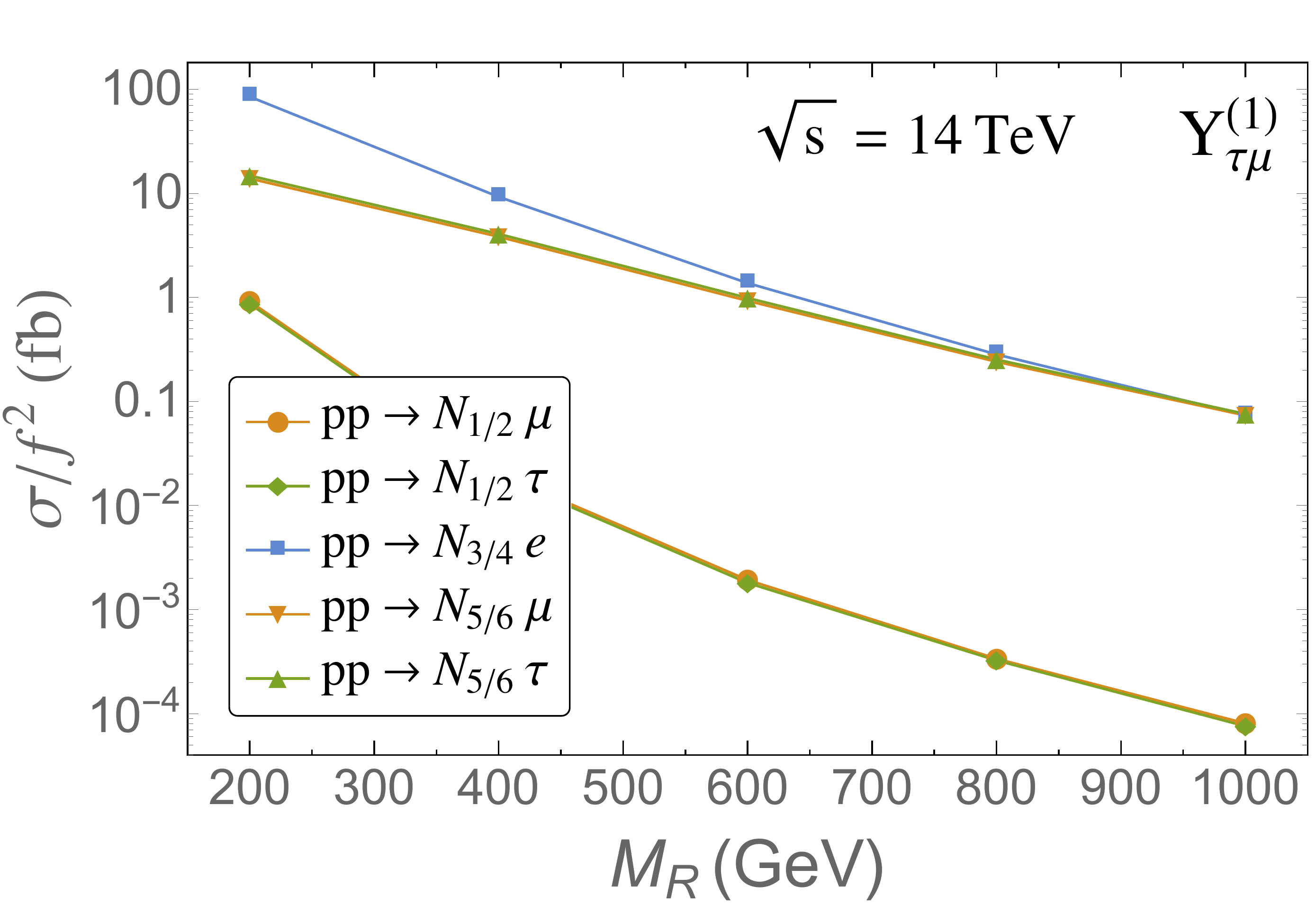}
\includegraphics[width=.95\columnwidth]{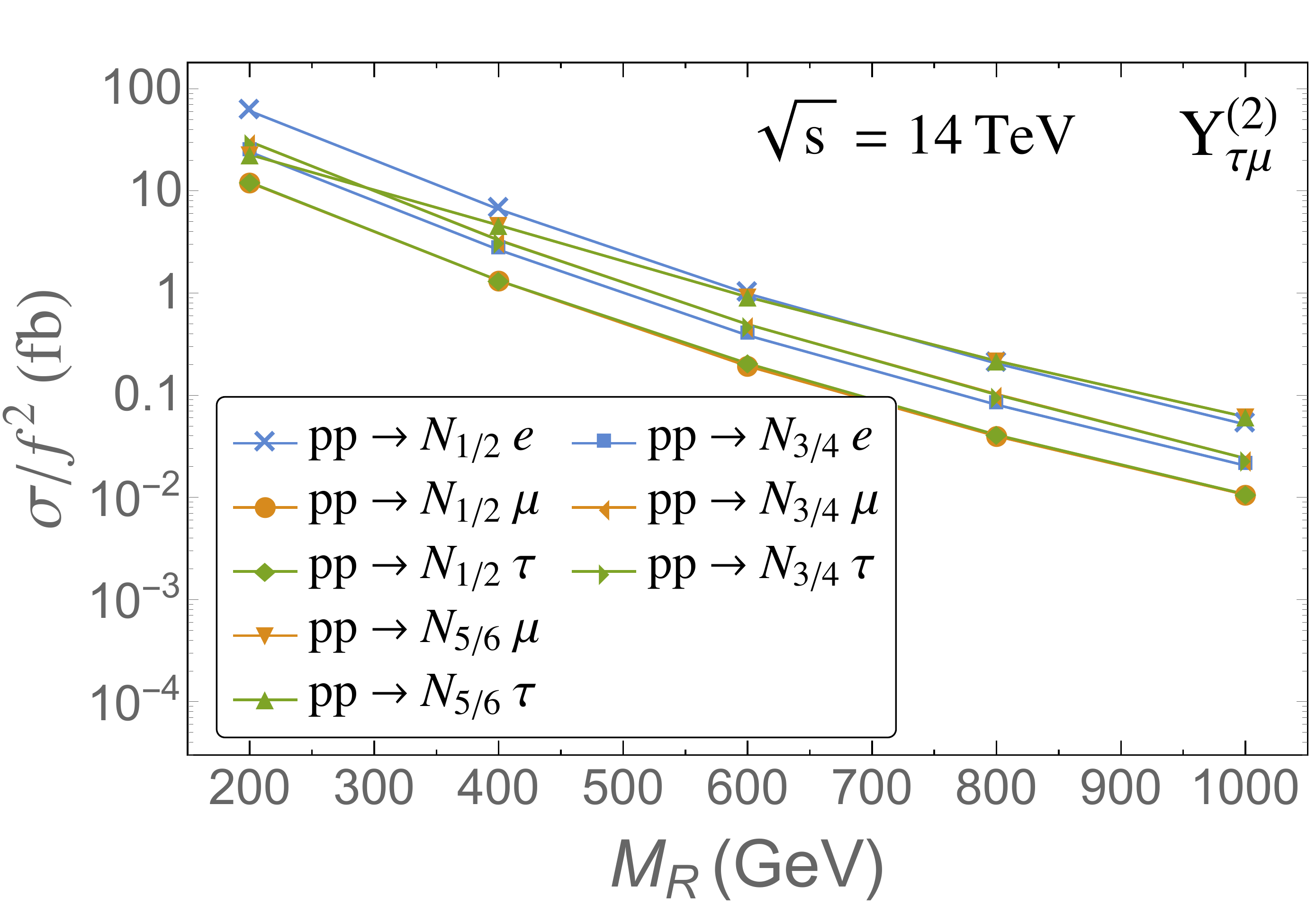}
\includegraphics[width=.95\columnwidth]{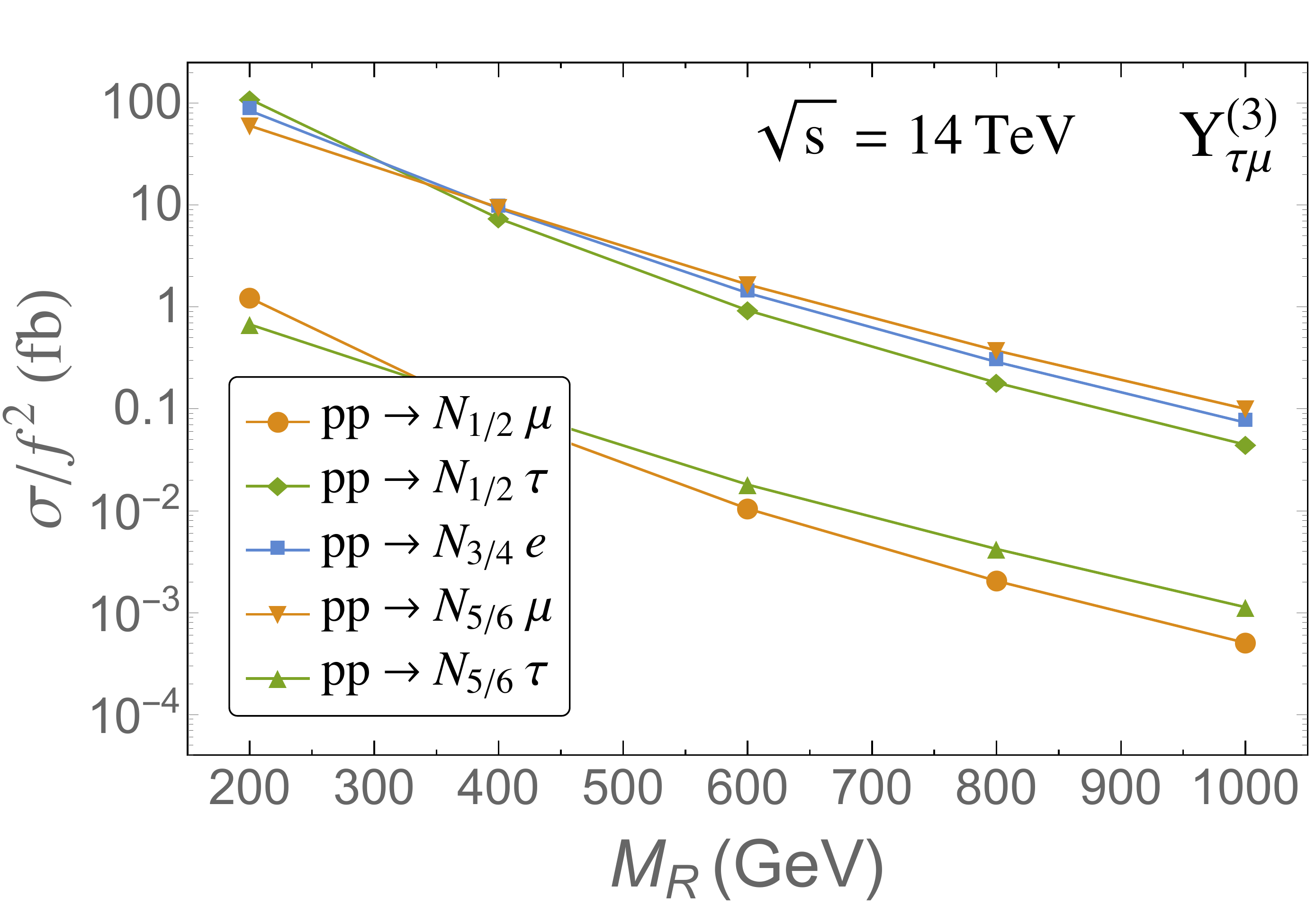}
\figcaption{Heavy neutrino DY-production  at the LHC for the three scenarios.   The cross sections are normalized by the Yukawa coupling intensity $f$ in (\ref{Ytmmax}). Processes not shown are negligible.}
\label{production}}
We can also learn that the flavor of the associated charged lepton is different depending on the heavy neutrino produced and the scenario considered, and that this pattern can be understood looking at the mixing in fig.\ref{mixings}.
Thus, $N_{3/4}$ are mainly electronic neutrinos in the first and  third scenarios, therefore, they are basically produced exclusively with electrons. $N_{1/2}$ are equally produced with muons and taus in the first scenario, dominantly produced with electrons in the second scenario, and mainly produced with taus in the third scenario. $N_{5/6}$ are equally produced with muons and taus in the first and second scenarios,  and mainly produced with muons in the third scenario.

Once the heavy neutrinos are produced, they will decay inside the detector. 
As mentioned above, in the limit $M_R\gg m_D$ the heavy neutrino masses are close to $M_R$, with small differences of $\mathcal O(m_D^2M_R^{-1})$ between the different pseudo-Dirac pairs and, therefore, assuming they are practically degenerate,  their decay into each other should be suppressed, with the dominant channels being $N_j\to Z\nu_i, H\nu_i, W^\pm l_i^\mp$. The relevant decays into $W^+l_i^-$ or  $W^-l_i^+$ have a partial width given by:
\begin{equation}
\Gamma_{N_j \to W l_i}=\frac{\sqrt{\big(M_j^2-m_i^2-m_W^2\big)^2-4 m_i^2 m_W^2}}{16\pi M_j^3}~\big| \overline {F_W} \big|^2,
\end{equation}
with
\begin{align}
\big| \overline {F_{W}} \big|^2 &=\frac{g^2}{4M_W^2} \big|B_{l_iN_j}\big|^2 \nonumber\\ 
&  \times\Big\{ \big(M_j^2-m_i^2\big)^2+M_W^2 (M_j^2+m_i^2)-2M_W^4\Big\}\,.
\end{align}
We have also seen that, in agreement with \cite{Atre:2009rg,OkadaTrilepton},  in the large $M_j$ limit, the heavy neutrinos decay with equal branching ratios to all the bosons in all the scenarios if we sum over all the flavors,  namely,
\begin{align}
{\rm BR}(N_j \to h \nu)&={\rm BR}(N_j \to Z \nu) =\nonumber\\ 
{\rm BR}(N_j \to W^+ l^-)& ={\rm BR}(N_j \to W^- l^+)=25\%\,.
\end{align} 
It is, however, interesting to study the rich flavor structure of the decay products, which depends on the decaying heavy neutrino  and the scenario we are considering.
Like in the production, the flavor preference of the decays to $W^\pm l^\mp$,  which are the ones relevant to this study and are given in fig.\ref{FDs}, also follows the same pattern as in fig.\ref{mixings}.
Therefore, we can expect the production and decay of the heavy neutrinos to lead to exotic $\mu\tau jj$ events with no missing energy and $M_{jj}\sim M_W$, with $M_{jj}$ the invariant mass of the two jets.

Using the narrow width approximation, the total cross section of the exotic events we are interested in is given by:
\begin{align}
\sigma(pp\to \mu\tau j j) &=\nonumber\\
\sum_{i=1}^{6}  &
\Big\{ 
\sigma(pp \to N_i \mu^\pm) {\rm BR}(N_i\to W^\pm \tau^\mp)\nonumber\\
& +\sigma(pp \to N_i \tau^\pm) {\rm BR}(N_i\to W^\pm \mu^\mp)  \Big\} \nonumber\\
&\times {\rm BR}(W^\pm\to jj).
\end{align}

\myfigure{
\includegraphics[width=\columnwidth]{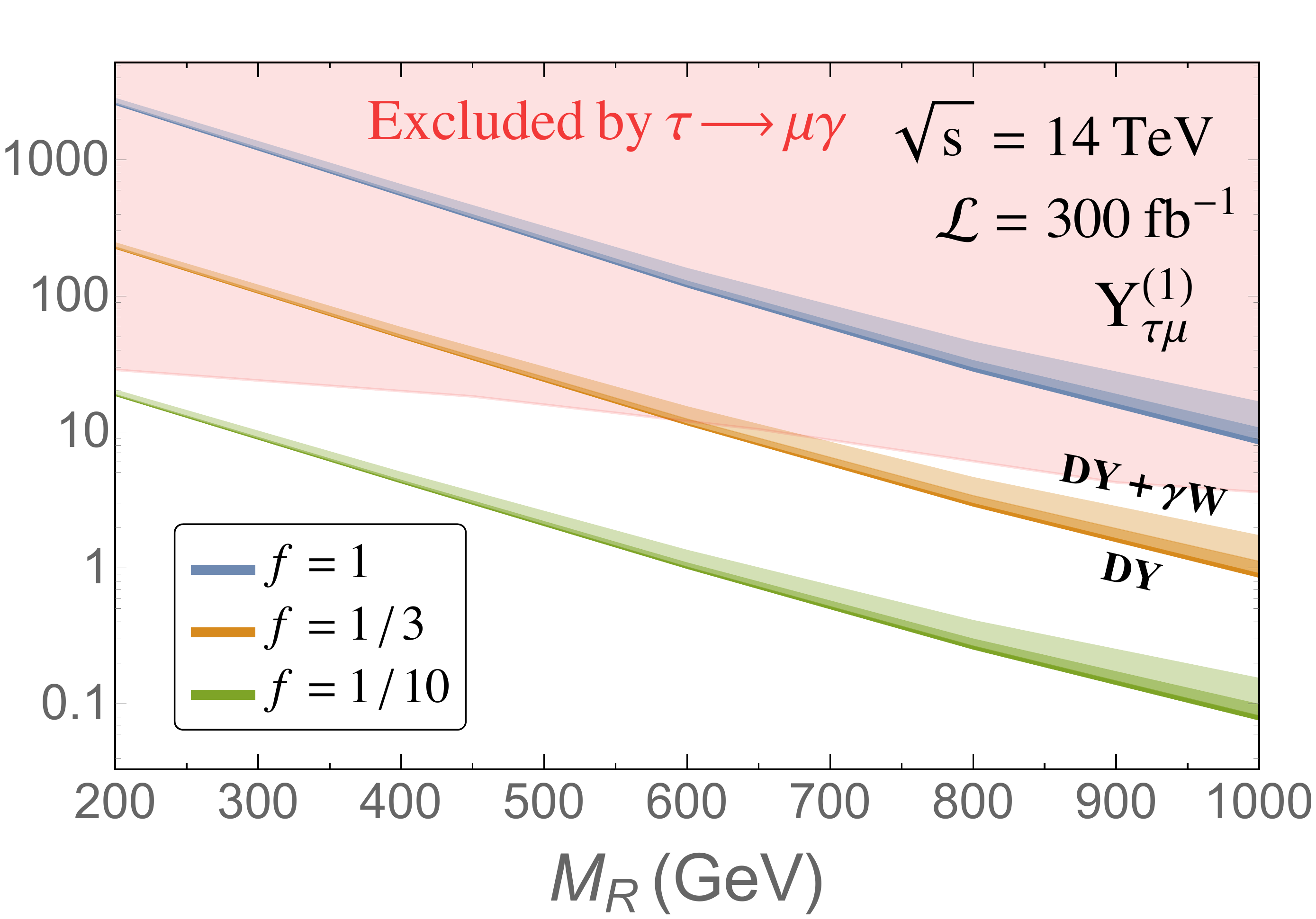}
\includegraphics[width=\columnwidth]{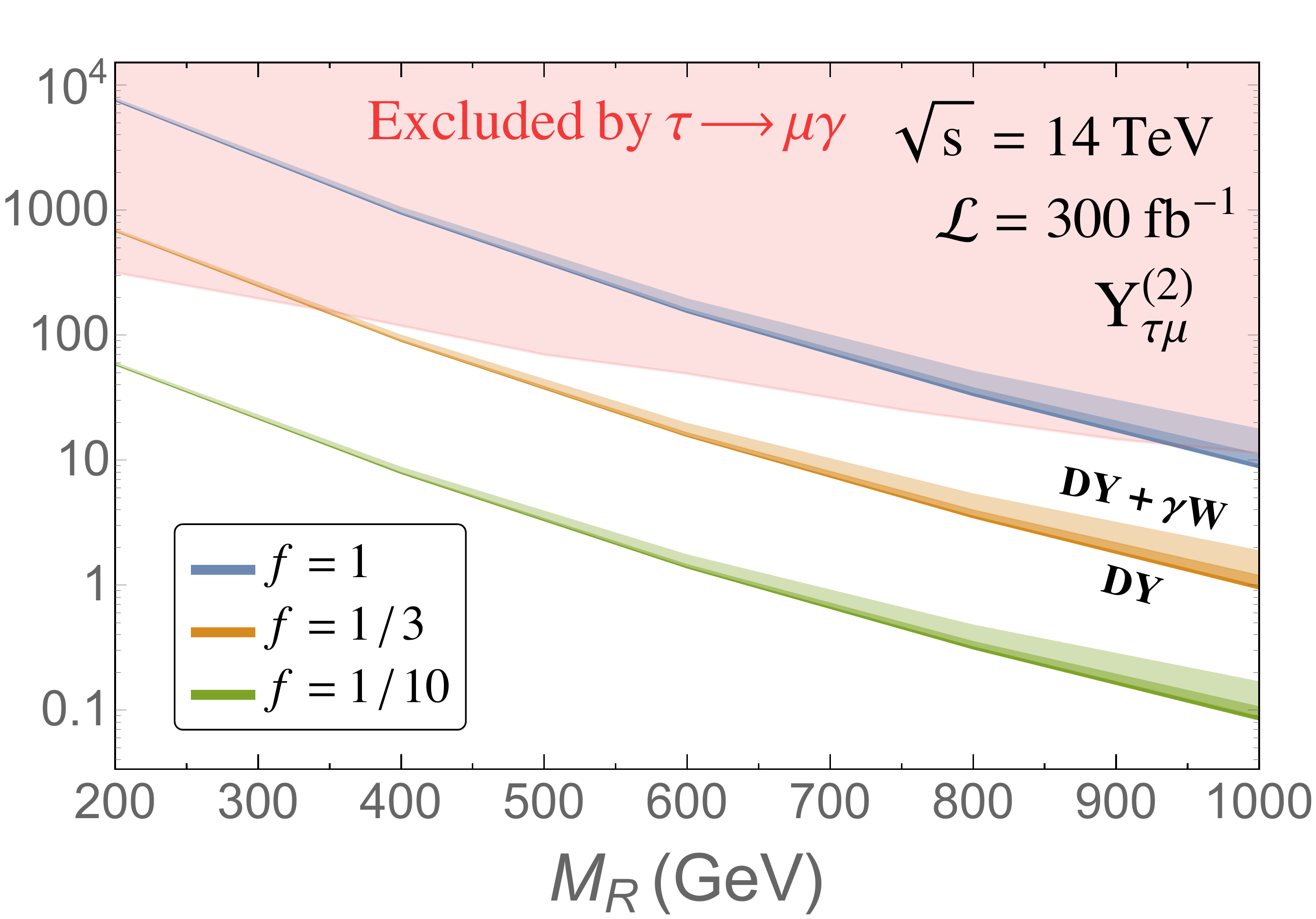}
\includegraphics[width=\columnwidth]{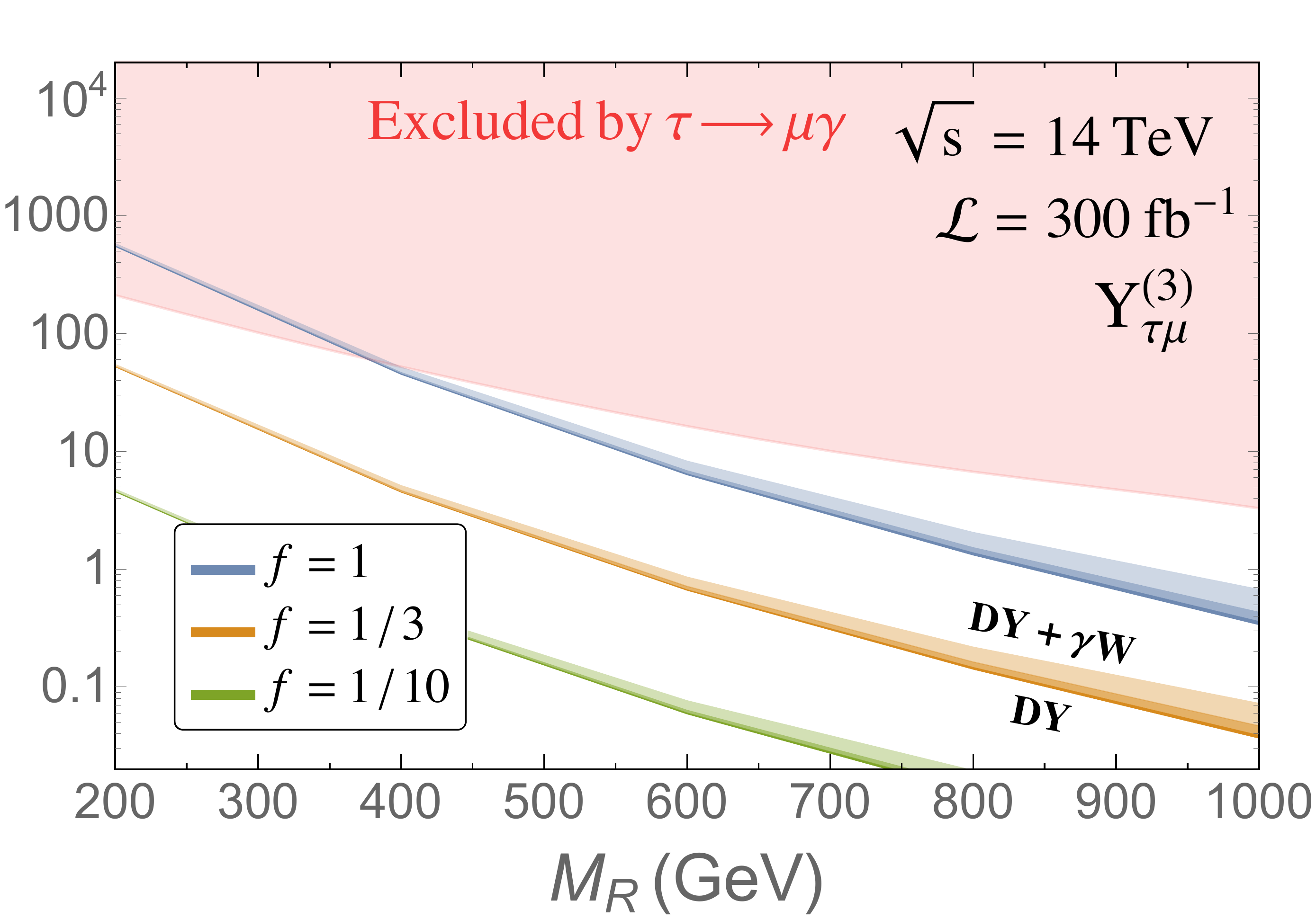}
\figcaption{Number of exotic $\mu\tau j j$ events at the LHC for the three scenarios and for three values of $f$. For each $f$, the bottom solid line is the prediction of $\mu\tau j j$ events from DY  and the upper lines on top of each of the three shadowed regions are the predictions after adding the $\mu\tau jj$ events from $\gamma W$, for $p_T^{\rm max}=10,20,40$ GeV from bottom to top, respectively. The $p_T^{\rm max}$ cut is only applied to the two extra jets in the events from $\gamma W$. The upper red shadowed areas are excluded by  $\tau\to\mu\gamma$.}
\label{events}}

Figure \ref{events} shows the expected number of exotic events $\mu\tau j j$ at the LHC for an integrated luminosity of $\mathcal L~=~300~ {\rm fb}^{-1}$ at $\sqrt s=14$ TeV.
The lower solid lines for each choice of $f$ are the number of events considering only the DY-production. 

Moreover, $\gamma W$ fusion processes can also contribute to this kind of exotic events if the $p_T$ of the extra jets are below a maximum value $p_T^{\rm max}$ and, therefore, they can be considered as soft or collinear jets which can escape detection. In this case the predicted total number of exotic events are the sum of the events produced by DY and  $\gamma W$ channels.
These total contributions for different values of $p_T^{\rm max}=10,20$ and $40$ GeV are shown as  the border lines on top of the shadowed areas with gradual decreasing intensity above each solid line. In addition we have included in the plots red shadowed areas that represent the regions excluded by the experimental upper bound on BR($\tau\to\mu\gamma$).
We can see that, after considering all the constraints, the three scenarios lead to an  interesting number of $\mathcal O(10-100)$ total $\mu\tau jj$ exotic events for the range of $M_R$ studied here of [200 GeV, 1 TeV]. 

The SM backgrounds for events with two leptons of different flavor have been studied in ref.\cite{Aad:2015pfa}. 
However, a high efficiency in the $\tau$-tagging and a good reconstruction of the $W$ boson invariant mass from the two leading jets would help in reducing the background. In that case, the main background would come from processes with photons or jets misidentified as leptons, mainly from $W/Z+\gamma^*$, $W/Z$ + jets and multijet events with at least four jets with one of them misidentified as a muon and another one as a tau; and from  $Z/\gamma^*+{\rm jets} \to\mu^+\mu^-~+$ jets if one of the muons is misidentified as a $\tau$ candidate.
Nevertheless, a dedicated background study for these particular $\mu\tau jj$ exotic events  is beyond the scope of this letter and it will be done in a future work. 

\section{Conclusions}
\label{conclusions}

In this letter we have proposed a new interesting way to study the production and decay of the heavy neutrinos of the ISS in connection with LFV. 
We have presented the computation of the predicted number of exotic $\mu\tau jj$ events which can be produced in these ISS scenarios with large LFV by the production of heavy pseudo-Dirac neutrinos together with a lepton of flavor $\ell$, both via DY and $\gamma W$ fusion processes, and their subsequent decay into $W$ plus a lepton of different flavor.
We have concluded that, for the three benchmark scenarios studied here, a number of $\mathcal O(10-100)$  total $\mu\tau jj$ exotic events without missing energy can be produced at this ongoing run of the LHC, for values of $M_R$ from 200 GeV to 1 TeV. 
Similarly, other rare processes like $\tau e jj$ or $\mu e jj$ could be produced within other ISS scenarios with large LFV, although for the latter ones, while being interesting since they could provide in addition observable CP asymmetries~\cite{Bray:2007ru}, the number of events would be strongly limited by the $\mu \to e \gamma$ upper bound.
Of course, a more realistic study of these exotic events, including detector simulation, together with a full background study should be done in order to reach a definitive conclusion, but this is beyond the scope of this letter and will be addressed in a future work.

\section*{Acknowledgments}

This work is supported by the European Union Grant No. FP7 ITN
INVISIBLES (Marie Curie Actions, Grant No. PITN- GA-2011- 289442), by the CICYT through Grant No. FPA2012-31880,  
by the Spanish Consolider-Ingenio 2010 Programme CPAN (Grant No. CSD2007-00042), 
and by the Spanish MINECO's ``Centro de Excelencia Severo Ochoa'' Programme under Grant No. SEV-2012-0249.
E.A. is financially supported by the Spanish DGIID-DGA Grant No. 2013-E24/2 and the Spanish MICINN Grants No. FPA2012-35453 and No. CPAN-CSD2007-00042.
X.M. is supported through the FPU Grant No. AP-2012-6708. C.W. receives financial support as an International Research Fellow of the Japan Society for the Promotion of Science.

\bibliographystyle{unsrt}

\end{multicols}

\end{document}